# *In-situ* nanoscale transport measurements on monoatomic metal films by low-temperature scanning tunneling potentiometry

Masayuki Hamada, Masahiro Haze*, Junya Okazaki, and Yukio Hasegawa

*Institute for solid state physics, The University of Tokyo*

*Kashiwanoha 5-1-5, Kashiwa, Chiba 277-8581, Japan*

*haze@issp.u-tokyo.ac.jp

## Abstract

Investigation of transport properties is fundamental for characterizing electronic properties and phase transitions. However, most of the transport measurements on conductive layers have been performed at macroscopic scales, and thus the development of microscopic methods to measure transport is important. Scanning tunneling potentiometry (STP) is a powerful tool for investigating surface conductivity at nano-scale spatial resolutions. However, it is still challenging to conduct STP studies at low temperatures and most of the low-temperature studies were performed on samples that were prepared *ex-situ*. In this study, we developed a low-temperature STP and demonstrated its performance on monoatomic metal films formed on Si(111) substrates that were prepared *in-situ*. Stable operation at low temperatures enables us to extract the electrochemical potential originating from the surface transport by canceling out the potential due to thermal differences and artifacts arising from the nonlinearity of the density of states (DOS). We also formulated the nonlinear-DOS artifact and confirmed it by comparing with the nonlinearity obtained by scanning tunneling spectroscopy.

# 1. Introduction

Two-dimensional (2D) transport, which includes surface transport, is unique compared to three-dimensional one [1–3]. Point defects, such as atomic defects and vacancies, and one-dimensional defects, such as steps and domain boundaries, have significant influences on the 2D transport properties [4,5]. However, the role of such local disorders has not been thoroughly investigated because of the lack of microscopic techniques that enable us to directly probe the transport around them. Moreover, at low temperatures, quantum phenomena, such as localization, charge density waves and superconductivity, are expected. Therefore, it is important to develop a system that can reveal microscopic transport properties at low temperatures.

To investigate local surface conductivity, scanning tunneling potentiometry (STP) has been proposed [6], which is a probe technique based on scanning tunneling microscopy (STM). By applying a lateral voltage across the sample surface, a spatial gradient of electrochemical potential is generated, which is closely related to the spatial variation in the local conductivity. In STP, an STM tip is used as a probe for the local electrochemical potential through a balance to achieve zero tunneling current. The comparison of the electrochemical potential image with the topographic one by STM and spectroscopic information by tunneling spectroscopy allows us to investigate the correlation between the local conductivity and the structural/spectral information.

To characterize the surface transport properties, the multiprobe technique has been used [1]. The method allows us to investigate surface transport properties with high sensitivity, but the spatial resolution is limited by the probe spacing, which is at least on the micrometer scale. Therefore, the influence of local defects distributed in smaller scales should be averaged. On the other hand, STP exhibits spatial resolution comparable with STM. Since the first development of STP, the method has been well established and has revealed the surface conductivity of various metallic phases on semiconductors [4,5,7–9], 2D materials [3,10–14], and topological insulators [15–17]. Although a few STP studies have been conducted at cryogenic temperatures [10–14], their samples are mostly prepared ex-situ. Therefore, it is important to set up a system with which one can perform STM in-situ on samples that otherwise might be contaminated and damaged.

In the STP measurements, a potential gradient is generated in a sample. As a consequence, the tunneling current between a tip and a sample contains both topographic and potential information. The separation of these signals is one of the most important issues for the STP operation. Methods of the separation developed so far are roughly divided into two types: dc (direct current) operation and ac (alternative current) operation. In the dc operation, the topographic information is obtained from the conventional STM feedback loop, and subsequently, the potential information is obtained. Because topographic and potential information are obtained separately in the time domain, this operation is relatively time-consuming. On the other hand, in the ac operation, the topographic and potential information are obtained simultaneously and are separated in the frequency domain. Whereas the measurement is faster, artifacts often appear because of the non-linearity of the local density of states (LDOS), as described later. Such an artifact can be canceled by comparing the signals obtained in the opposite current directions through a sample [18]. However, for further understanding, a quantitative discussion is necessary using a sample that has an inhomogeneous LDOS.

Here, we report the development of our STP setup and the demonstration of the measurements at low temperatures (LT-STP). First, to demonstrate the performance of our LT-STP, we measured the surface potential of the polycrystalline Au film, which was prepared *ex-situ* [19]. Subsequently, we demonstrated STP on atomically flat monolayer films formed on semiconducting substrates, which were prepared *in-situ*. We employed a Pb monolayer formed on Si(111) as a sample because it is one of the most investigated metallic phases on a semiconducting substrate [20]. We fabricated a striped incommensurate (SIC) phase that includes √3×√3 domains, their domain boundaries, and 2 monolayers region. Because of the presence of various phases, LDOS are inhomogeneous and thus the artifact due to nonlinear DOS appears in the STP images. Moreover, the system has attracted significant attention in terms of superconductivity and its strong relation with disorders [21–23].

## 2. Methods

We developed an LT-STP setup by incorporating an STP circuit into a commercial low-temperature UHV-STM (Unisoku USM-1300S) and a commercial STM feedback system (RHK R9 controller). The system is cooled by liquid helium and the base temperature, which is measured by the Cernox sensor located close to the sample holder, is 7 K. Our STP circuit is based on a previous report [8], and its details are found elsewhere [19]. A schematic of the STP circuit is shown in Fig. 1. To create a lateral current flowing parallel to the sample surface, the voltage $V_{lat}$ = $V_1 - V_2$ is applied between the two electrodes across the sample surface. The overall electrochemical potential distribution is illustrated by the dashed line in Figure 1. Due to the applied voltage, the potential under the tip depends on its position on the surface. The potential information is obtained by the STP feedback loop, which controls an offset of the lateral voltage to set the dc component ($I_{dc}$) of the tunneling current ($I_t$) to zero. The output voltage of the STP feedback multiplied by -1, which we call STP signal, provides a potential on the sample just below the tip.

To obtain a topographic STM image, the ac voltage ($V_{ac}$) is superimposed on the offset voltage. Thus, the total tunneling current $I_t$ is the sum of the ac and dc components: $I_t = I_{ac} + I_{dc}$. As described above, $I_{dc}$ is set to zero by the STP feedback, whereas $I_{ac}$ is not. Since the frequency of $I_{ac}$ is higher than the cutoff of the STM feedback, the rectified signal of $I_{ac}$ is time-averaged and loaded as an input signal for the STM feedback. In this operation, which is called the ac-STP operation, the obtained STP signal includes not only the electrochemical potential $\mu_{ecp}$, but also the voltage due to the nonlinearity of DOS $V_{DOS}$.

$V_{DOS}$ arises from the nonlinearity of the DOS at $V = 0$, as shown schematically in Fig. 2(a). In our STP feedback loop, ac voltage with amplitude $V_{ac}$ is applied, and $V_{STP}$ is set to make the time-averaged tunneling current zero. When the current ($I(V)$) varies linearly with the sample bias voltage around $V = 0$, which corresponds to the case where the second derivative at $V = 0$ is zero, the center of $V_{ac}$ is set to zero by the STP feedback. However, when the second derivative at $V = 0$ is not zero, the center of the $V_{ac}$ is not zero. Figure 2(b) shows a case in which the second derivative is positive. To make the time-averaged tunneling current zero, the STP feedback should apply a negative voltage to compensate for the larger tunneling current at the positive side of the bias voltage; thus, the center of the $V_{ac}$ is shifted to the negative side.

To quantitatively evaluate the shifted voltage, which corresponds to the $V_{DOS}$, we consider the time-averaged current $\tilde{I}$, which is described by

$$\tilde{I} = \frac{1}{T}\int_0^T [I(V_{DOS} + V_{ac}\sin\omega t)]\,dt. \quad (2)$$

We then apply the Taylor expansion around $V = V_{DOS}$ to the integrands up to the second order.

$$\tilde{I} = \frac{1}{T}\int_0^T \left[I(V_{DOS}) + V_{ac}\sin\omega t \left.\frac{dI}{dV}\right|_{V=V_{DOS}} + \frac{1}{2}V_{ac}^2\sin^2\omega t \left.\frac{d^2I}{dV^2}\right|_{V=V_{DOS}}\right]dt. \quad (3)$$

$$\tilde{I} = \frac{1}{T}\int_0^T \left[I(V_{DOS}) + V_{ac}\sin\omega t \left.\frac{dI}{dV}\right|_{V=V_{DOS}} + \frac{1}{4}V_{ac}^2(1-\cos 2\omega t) \left.\frac{d^2I}{dV^2}\right|_{V=V_{DOS}}\right]dt. \quad (4)$$

By the integration, the terms containing sin$\omega t$ and cos2$\omega t$ are zero. As described above, by the STP feedback, $\tilde{I}$ is set to 0. Therefore,

$$I(V_{DOS}) = -\frac{1}{4}V_{ac}^2 \left.\frac{d^2I}{dV^2}\right|_{V=V_{DOS}}. \quad (5)$$

Here, the Taylor expansion is again applied to both sides around $V = 0$,

$$I(0) + V_{DOS}\left.\frac{dI}{dV}\right|_{V=0} = -\frac{1}{4}V_{ac}^2 \left.\frac{d^2I}{dV^2}\right|_{V=0} - \frac{1}{4}V_{DOS}V_{ac}^2 \left.\frac{d^3I}{dV^3}\right|_{V=0}. \quad (6)$$

Since $I(0)$ is zero and we assume that the third-order derivative is negligibly small, $V_{\text{DOS}}$ is derived as follows,

$$V_{\text{DOS}} = -\frac{1}{4}V_{\text{ac}}^2 \frac{d^2I/dV^2}{dI/dV}\bigg|_{V=0}. \quad (7)$$

In addition to $V_{\text{DOS}}$, the Seebeck voltage $V_{\text{th}}$ may appear in the STP signal. $V_{\text{th}}$ is generated by the temperature gradient between the tip and the sample as follows [18,24,25].

$$V_{\text{th}} = \frac{\pi^2 k_B^2}{6e^2}(T_\text{T}^2 - T_\text{S}^2)\frac{d^2I/dV^2}{dI/dV}\bigg|_{V=0}, \quad (8)$$

where $T_\text{T}$ and $T_\text{S}$ are the temperatures of the tip and sample, respectively, $k_B$ is the Boltzmann constant, and $e$ is the elementary charge. Because $V_{\text{DOS}}$ is due to the nonlinearity of the DOS of both the sample and tip, it is independent of $V_{\text{lat}}$. $V_{\text{th}}$ is a thermal effect and thus may be affected by the amount of $V_{\text{lat}}$. However, the Joule heating does not depend on the $V_{\text{lat}}$ polarity. Considering the –1 multiplication, the STP signal $V_{\text{STP}}$ is expressed as follows,

$$V_{\text{STP}}(V_{lat}) = \mu_{\text{ecp}}(V_{lat}) - V_{\text{th}}(|V_{\text{lat}}|) - V_{\text{DOS}}, \quad (8)$$

using the electrochemical potential of the sample $\mu_{\text{ecp}}$. Therefore, $\mu_{\text{ecp}}$ is extracted by subtraction as follows [18],

$$\mu_{\text{ecp}}(V_{\text{lat}}) = \frac{1}{2}\{V_{\text{STP}}(V_{\text{lat}}) - V_{\text{STP}}(-V_{\text{lat}})\}. \quad (9)$$

## 3. Polycrystalline Au films formed on SiO$_2$ substrate

### 3-1 Sample

To demonstrate the performance of our STP system, we performed LT-STP measurements on a polycrystalline Au thin film deposited by an ion coater on a SiO$_2$-coated (thickness: 200 nm) Si substrate. Prior to the deposition of the Au film, two rectangular Au electrodes were formed using a mask. Figure 3(a) shows a photo of a deposition stage with "mask A" for fabricating the electrodes, and Fig. 3(b) shows a schematic side view of (a). Figures 3(c) and 3(d) show a photo of the electrodes on a sample and a schematic side view of (c), respectively. The thickness of the electrode is ~40 nm and the gap distance between the two electrodes is ~2 mm. After the electrode fabrication, Au thin film for the STP measurements were deposited so that it bridges the two electrodes. Figure 3(e) shows a photo of a deposition stage with "mask B" for forming the Au thin film, and Fig. 3(f) shows a schematic side view of (e). Figures 3(g) and (h) show a photo of the sample after the formation of the Au thin film between the electrodes and a schematic side view of (g), respectively. The thickness of the Au thin film was ~5 nm

## 3-2 Results

Figure 4(a) shows a topographic STM image of the Au thin film. The granular structures of Au whose size is ~20 nm can be observed. The averaged roughness is ~0.7 nm. Figures 4(b) and 4(c) show the electrochemical potential (STP) images taken on the same area as (a). These two images were obtained by applying opposite lateral voltages ($V_{lat} = \pm 2.3$ V) between the two Au electrodes. In Fig. 4(b), the overall contrasts are tilted from left to right, and the slope is reversed in Fig. 4(c). The potential gradient reversal along the current directions indicates that the electrochemical potential was properly measured and imaged. The temperature during the measurement was 19 K, which is above the base temperature, because of the Joule heating by the lateral current. Here we note that the temperature is not of the sample temperature but of the STM system, as explained above. Figure 4 (d) is an image of (b) subtracted by (c), as described in Eq. (9). As discussed above, the subtracted image corresponds to the pure electrochemical potential image. In the potential image, abrupt potential drops are observed at some grain boundaries, indicating that these grain boundaries have electrical resistance. We here note that not all grain boundaries have such resistance. This potential image demonstrates that STP measurements can identify the presence and spatial distribution of local resistances, which cannot be resolved in topographic images. Figure 4(e) shows the cross-sectional plots of the topographic height and potential taken along the blue and red dashed lines in (a) and (d), respectively. An abrupt potential drop ($\Delta V$ = ~200 $\mu$V) is observed at the central grain boundary. The black dashed line in Fig. 4(e) indicates the curve fitted using the logistic equation, which describes the smeared step function.

$$f(x) = A + \frac{B}{1+\exp(-(x-x_0)/k)}. \quad (10)$$

Here, $x$ is the position, $x_0$ is the center of the step function, $A$ is the offset, and $B$ is the height of the step function. $k$ is the smearing factor. From the fitting, we obtain $A$ = 56 $\mu$V, $B$ = 227 $\mu$V, $x_0$ = 154.6 nm, and $k$ = 0.4 nm. While it has to be careful to discuss the spatial resolution because of the artifacts caused by an STM tip shape [26,27], we estimate from the value of $k$ that the spatial resolution of the potential measurement is in nm scales. The potential is almost flat in the other area, suggesting that the global resistance is dominated by the grain boundaries, which are essentially the same as the previous result performed at room temperature [19]. Since the standard deviation of the potential in the flat area is ~20 $\mu$V, our STP system has a potential resolution in a few tens $\mu$V scale. Therefore, we conclude that our LT-STM works well with high spatial and potential resolution.

# 4. Pb monolayer formed on Si(111) substrate

## 4-1. Sample

To perform LT-STP measurements on surface reconstructed structures, we fabricated one of the metallic phases of Pb monolayers on Si(111) substrate: the striped incommensurate (SIC) phase with a coverage of ~4/3 monolayers. It has been known that this phase is metallic at our measurement temperature (7-20 K) and shows a superconducting transition at 1.51 K [23]. To measure the pure conductivity of the overlayers, we used a non-doped Si(111) substrate, which has a resistivity more than 1 k$\Omega$·cm at room temperature. Since it becomes an insulator at low temperatures, the conductivity through the substrate can be ignored.

To prepare a clean surface on the Si substrate, high temperature flashing (~1200 °C) is required. Therefore, we used Ta (tantalum) electrodes instead of Au because of its high melting temperature (3020 °C) and chemical inertness. We employed an ion-beam sputtering deposition method through a shadow mask to fabricate the electrodes on the substrate. Figures 5(a) and 5(b) show a photo and a schematic side view of the deposition stage and the mask for fabricating electrodes. The size of the Si substrates was 2 × 11 mm$^2$. Figure 5(c) is a photo of the sample after forming the electrodes with a schematic side view of (c) shown in Fig. 5(d). The thickness of the electrodes is ~380 nm and the gap distance between the two electrodes, where a bare Si surface is exposed, is 2 mm. An atomically clean surface can be fabricated in this gapped area after several high temperature flashes without damaging the electrodes.

After loading in the ultra-high vacuum (UHV) STM chamber, the Si(111) substrate with the Ta electrodes was degassed at 600 °C under the UHV condition for more than 6 hours. Subsequently, the substrate was flashed up to 1200 °C several times to obtain a clean Si(111)-7×7 reconstructed surface. Pb was deposited from an e-beam evaporator onto the substrate maintained at room temperature. To form the SIC phase, the sample was annealed at 350 °C for 3 minutes after the deposition. After cooling to a low temperature, the resistance across the sample is ~1 k$\Omega$ between the electrodes. Because the resistance without the Pb overlayer is more than 1 G$\Omega$ at low temperatures, the current flows only through the overlayer.

## 4-2. results

Figure 6 (a) shows a topographic STM image showing steps, terraces, and islands structures. The inset shows a zoomed topographic image taken in a terrace, which exhibits the SIC phase [21].

Figures 6(b) and 6(c) show the surface potential images taken on the same area during the current flowing from left to right and vice versa, respectively. In the potential images, the contrast changes gradually along each current direction. However, in addition, contrasts that do not depend on the lateral current direction and thus are probably not related to the electrochemical potential are observed, *e.g.,* on the islands and at the domain boundaries of the SIC phase. To extract an electrochemical potential image $\mu_{\text{ecp}}$, we subtracted the image taken with the opposite lateral current (Fig. 6(c)) from the one with the forward lateral current (Fig. 6(b)), as described in the section on the Au films and Eq. (9). The subtracted image is shown in Fig. 6(d). Contrasts related to the islands and domain boundaries mostly disappeared so that we conclude that these are due to $V_{\text{th}}$ and/or $V_{\text{DOS}}$ (See Eqs. (8) and (9)), which will be discussed later in detail.

To investigate the local variation of the electrochemical potential, we obtained the cross-sectional profiles along the blue and red dashed lines in Figs. 6(a) and 6(d), as shown in Fig. 6(e). In the profile of the potential, a clear jump was not observed at the step edge, indicating that the resistance across the step is small compared to that of the terraces. In the case of Si(111)-√3×√3 Ag [5,7,9] and Si(111)-7×7 surfaces [4], the resistance of the steps is significantly high and dominant for their net resistance. We found that the SIC phase is the opposite case; the terrace resistance is dominant. From the linear fitting, we obtained that the slope is 1.2 $\mu$V/nm. Because the sample size is about 2 mm and the applied lateral voltage is 3.0 V, the global potential slope is estimated to be 1.5 $\mu$V/nm, which coincides well with that estimated from the STP. The small deviation is most probably due to the contact resistance between the Ta electrodes and the Pb film. The estimated contact resistance is 174 Ω, which is 1/4 the net resistance of our film.

Although the contrasts due to $V_{\text{DOS}}$ can be canceled by the subtraction, it is also important to evaluate $V_{\text{DOS}}$ from the d$I$/d$V$ spectra. We compared the d$I$/d$V$ spectra and STP signals taken on an island and a terrace. Figures 7(a) and 7(b) show the topographic and STP images taken around an island. In these measurements, the lateral voltage was not applied and therefore the obtained signal is only due to $V_{\text{DOS}}$. From the cross-sectional profile shown in Fig. 7(c), the signal difference between the terrace and the island is ~1700 $\mu$V. On the other hand, comparing the d$^2I$/d$V^2$ spectra taken on the two sites (Fig. 7(e)), which are numerical derivatives of the measured d$I$/d$V$ spectra shown in Fig. 7(d), we found that the one taken on the island shows a positive d$^2I$/d$V^2$ at V = 0 whereas the one taken on the terrace is almost zero. This agrees with the origin of $V_{\text{DOS}}$ (See Eqs. (7) and (8)); $V_{\text{DOS}}$ is higher on the island than on the terrace. From Figs. 7(d) and 7(e), we calculated $\left.\frac{\text{d}^2I/\text{d}V^2}{\text{d}I/\text{d}V}\right|_{V_{\text{dc}}=0}$ ~ 6.6 /V for the island and $\left.\frac{\text{d}^2I/\text{d}V^2}{\text{d}I/\text{d}V}\right|_{V_{\text{dc}}=0}$ ~ -0.18 /V for the terrace. We expect the $V_{\text{DOS}}$ difference is ~1200 $\mu$V for $V_{\text{ac}}$ = 25 mV from eq.(7), which is consistent with the experimentally measured difference in $V_{\text{DOS}}$.

## 5. Conclusion

In conclusion, we have developed an LT-STP system that works on metal monolayers formed on semiconducting substrates that were prepared *in-situ*. We first demonstrated the performance of LT-STP on a polycrystalline Au thin film formed on a $SiO_2$ substrate. We then performed LT-STP measurements on Si(111)-SIC Pb, which has atomically thin flat with multi-domain phases. In both cases, potential images exhibiting the spatial distribution of local resistances are obtained. Our stable low-temperature measurements allow us to extract a electrochemical potential image through the subtraction of two STP images taken under opposite current flows on the same area. In addition, we formulated an artifact arising from the nonlinearity of the DOS and confirmed it through a comparison with the results by scanning tunneling spectroscopy.

## ACKNOWLEDGMENTS

We thank laboratory of nanoscale quantum materials in the institute for solid state physics for the support performing the ion beam sputtering. This work is supported by the JSPS KAKENHI (Nos. JP24K01342, JP22K14598, JP22H00292, JP20K05319).

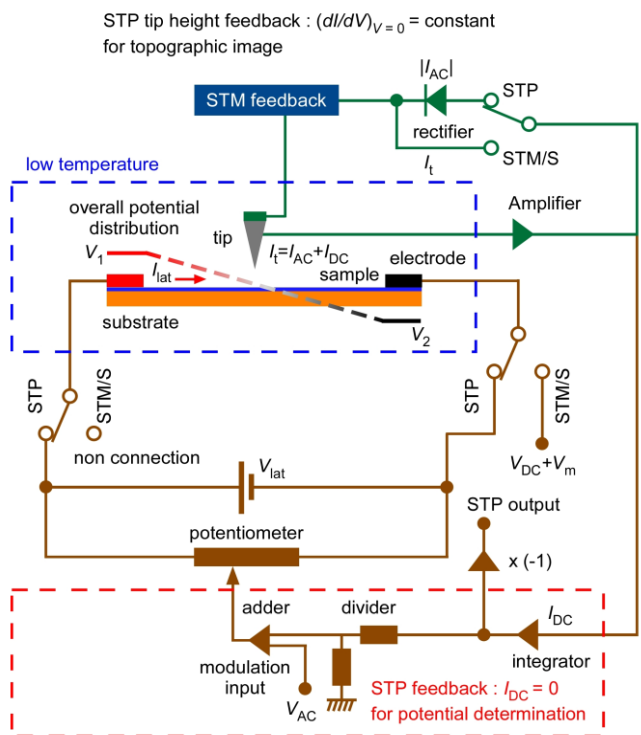

Fig. 1. Schematic diagram of our STP circuit.

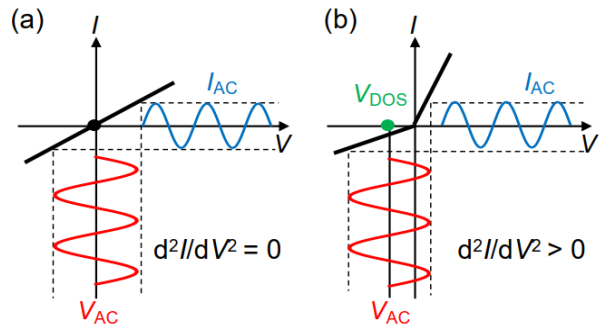

Fig. 2. Artifacts due to nonlinearity in the density of states. When $d^2I/dV^2 = 0$, $V_{DOS} = 0$ when $\tilde{I}$ is set to 0 by the STP feedback, as shown in (a). In contrast, when $d^2I/dV^2 > 0$, $V_{DOS} < 0$ as shown in (b).

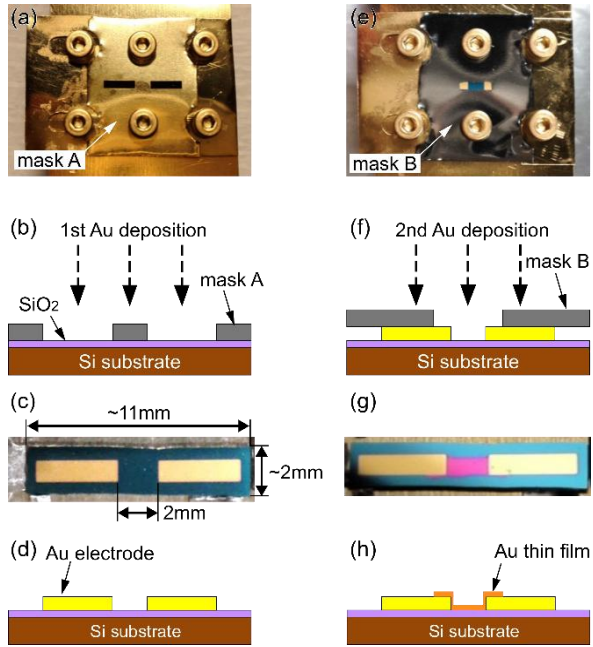

Fig. 3. (a) Photo of a deposition stage with "mask A" for fabricating the electrodes. (b) Schematic side view of (a). (c) Photo of a sample after forming the electrodes. (d) Schematic side view of (c). The thickness of the Au electrodes is ~40nm. (e) Photo of a deposition stage with "mask B" for forming Au thin films for the STP measurements. (f) Schematic side view of (e). (g) Photo of a sample after forming an Au thin film bridging the two electrodes. (h) Schematic side view of (g). The thickness of the second deposition is ~5 nm.

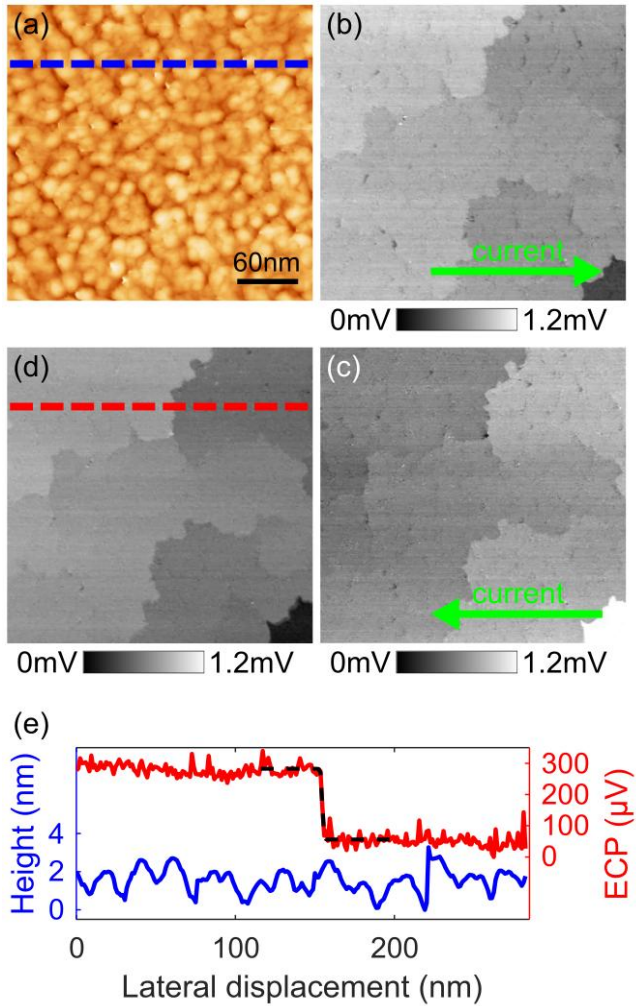

Fig. 4. (a) Topographic STM image taken on a Au thin film on the $SiO_2$-coated Si substrate. (b,c) electrochemical potential (STP) images taken on the same area during the lateral current flowing from left to right (from right to left), respectively. The images were taken at $T = 19$ K. The applied lateral voltage and current density flowing between the Au electrodes are $V_{lat} = \pm 2.3$ V and $J_{lat} = 7.4$ mA/mm, respectively. The amplitude of the modulation voltage $V_{ac} = 5$ mV with the frequency of 2 kHz. The rectified ac current $|I_{ac}|$ used for the STM feedback is set at 2 nA. (d) Subtracted image of (b) from (c). The values are divided by 2 after the subtraction. (e) Cross-sectional plots of the topographic height and potential taken along the dashed line drawn in (a) and (d), respectively. The black dashed line is a fitted curve using the logistic equation.

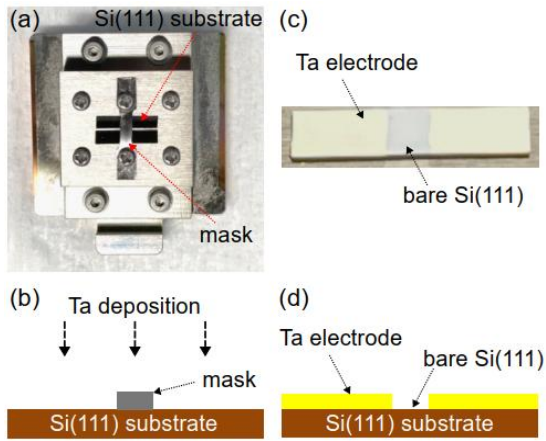

Fig. 5. (a) Photo of a deposition stage mounting two Si(111) wafer pieces and a mask for fabricating Ta electrodes. (b) Schematic side view of (a). (c) Photo of a sample after forming the electrodes. (d) Schematic side view of (c). The Ta deposition is performed by sputtering. The typical thickness of the electrodes is ~380 nm.

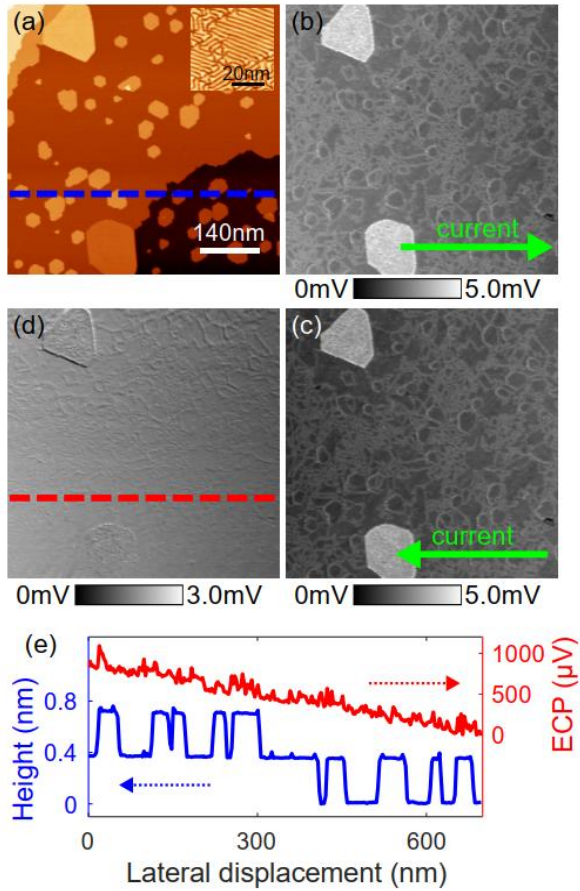

Fig. 6. (a) Topographic image of Pb-SIC formed on Si(111). (b,c) Potential images during current flow from right to left and vice versa, respectively. (d) Subtracted image of (c) from (b). The values are divided by 2 after the subtraction. The potential images were taken with the following parameters. $V_{lat} = \pm 3.0$V, $J_{lat} = 1.73$mA/mm, $V_{ac} = 15$mV, $|I_{ac}| = 200$pA, and $T = 18.9$K. (e) Cross-sectional profiles taken along the blue and red dashed lines in (a) and (d), respectively.

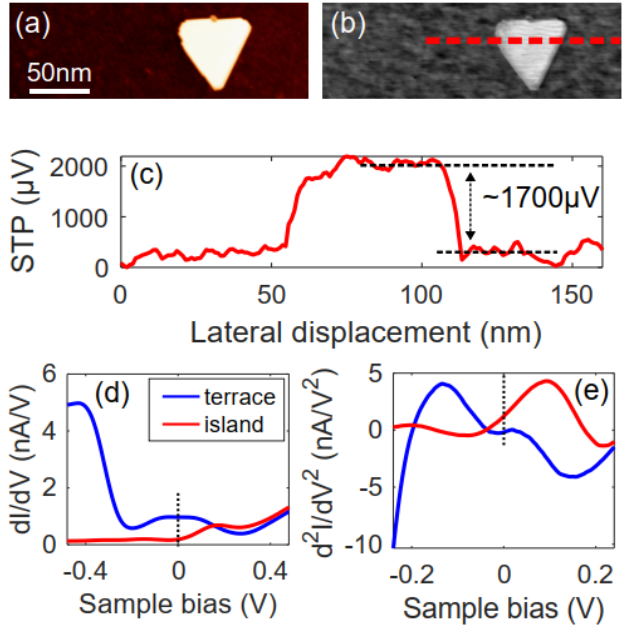

Fig. 7. (a,b) Topographic and STP images taken around an island. $V_{lat} = 0$ V, $J_{lat} = 0$ mA/mm, $V_{ac} = 25$ mV, $|I_{ac}| = 200$ pA, and $T = 7.2$ K. (c) Cross-sectional profile taken along the red dashed line in (b). (d) d$I$/d$V$ spectra taken on the terrace and the island (the tip was stabilized at $V_{dc} = 0.5$ V and $I_{dc} = 500$ pA). (e) d$^2I$/d$V^2$ spectra numerically differentiated from the spectra in (d).